\documentclass[preprint,prl,amsmath,amssymb]{revtex4-1}

\usepackage{graphicx}
\usepackage{mathptmx}

\usepackage{color}
\usepackage{ulem}

\begin{document}

\title{General approach to the understanding the electronic structure of graphene on metals}

\author{E. N. Voloshina\footnote{Corresponding author. E-mail: elena.voloshina@hu-berlin.de}}
\affiliation{\mbox{Institut f\"ur Chemie, Humboldt-Universit\"at zu Berlin, 10099 Berlin, Germany}}

\author{Yu. S. Dedkov\footnote{Corresponding author. E-mail: Yuriy.Dedkov@specs.com}}
\affiliation{\mbox{SPECS Surface Nano Analysis GmbH, Voltastra\ss e 5, 13355 Berlin, Germany}}

\date{\today}

\begin{abstract} 
This manuscript presents the general approach to the understanding of the connection between bonding mechanism and electronic structure of graphene on metals. To demonstrate its validity, two limiting cases of the ``weakly'' and ``strongly'' bonded graphene on Al(111) and Ni(111) are considered, where the Dirac cone is preserved or fully destroyed, respectively. Furthermore, the electronic structure, \textit{i.\,e.} doping level, hybridization effects, as well as a gap formation at the Dirac point of the intermediate system, graphene/Cu(111), is fully understood in the framework of the proposed approach. This work summarises the long-term debates regarding connection of the bonding strength and the valence band modification in the graphene/metal systems and paves a way for the effective control of the electronic states of graphene in the vicinity of the Fermi level.
\end{abstract}

\maketitle

The recent demonstration of the unique transport properties of graphene~\cite{Novoselov:2005,Zhang:2005}, a two dimensional allotrope form of carbon, opens a door in the world where strict 2D circuits built on the basis of graphene can be fabricated and used~\cite{Novoselov:2013hw}. However all these applications require that at some point the graphene-based device is contacted by metal. Considering graphene on metal one can expect, in general case, a strong modification of its valence band electronic structure. Presently, with respect to such changes, graphene on metal is described either as ``strongly'' or ``weakly'' bonded to metal. In the first case graphene is always $n$-doped and the overlap of the valence band states of graphene and metal at relatively short distance between them (in the range of $2$\,-\,$2.3$\,\AA) completely destroys the linear dispersion of the graphene $\pi$ states around the Fermi level ($E_F$)~\cite{Bertoni:2004,Karpan:2008,Weser:2011,Dedkov:2010a,Varykhalov:2009,Brugger:2009}. For the ``weakly'' interacting system (graphene can be $n$- or $p$-doped) a linear dispersion of $\pi$ states around $E_F$ is preserved~\cite{Dedkov:2001,Pletikosic:2009,Sutter:2009a,Varykhalov:2010a,Voloshina:2011NJP}.

Here it is worth to mention that DFT calculations with different functionals, in most cases giving very good agreement with experiment regarding the electronic structure, predict the bonding energy of graphene to metals in the range of $50$\,-\,$150$\,meV/C-atom with the higher value for the ``strongly'' bonded graphene~\cite{Voloshina:2012c}. This value is far below the lower limit of $\approx500$\,meV/atom which is usually taken for the description of the strong chemical adsorption on the metallic surfaces.

At the same time the recent angle-resolved photoemission experiments on graphene/Cu and graphene/Au show that despite of the linear behaviour of the graphene-derived $\pi$ states around $E_F$, there is a clear hybridization between graphene $\pi$ and metal $d$ valence band states~\cite{Shikin:2013fr}. These experiments also reveal the pronounced energy gap at the Dirac point ($E_D$)~\cite{Varykhalov:2010a}. The idea that this gap is due to the broken sublattice symmetry and that the width of the energy gap depends on the doping level of graphene is not supported by other experiments or calculations~\cite{Gruneis:2008,Voloshina:2011NJP,Petrovic:2013vz,Fedorov:2014aa}.

In this manuscript, basing on the analysis of a large amount of experimentally and computationally obtained band structures, we propose an universal model that allows to describe any graphene/metal system. All experimental observations as doping, hybridization of the valence band states of graphene and metal, and gap formation are considered and explained in the framework of this approach. The validity of our model is supported by the results of DFT calculations for several representative examples.

Let us consider graphene on the close-packed metallic substrate as shown in Fig.~\ref{structure_orbitals}. Without loss of generality (see Supplementary material~\cite{suppl:all} for details) we take a lattice-matched case of graphene on the (111) surface of metal. In such system two carbon atoms from different sublattices occupy inequivalent adsorption positions of the metallic surface. They are called $top$, $fcc$, and $hcp$ if carbon atom is placed either above M(S) atom or above the corresponding hollow site of the metallic slab, respectively [Fig.~\ref{structure_orbitals}(a)]. Usually in the ground-state structure one of the carbon atoms occupies $top$ position and the second one is placed above $fcc$ or $hcp$ position (see Ref.~\cite{Voloshina:2012c} and reference therein; Ref.~\cite{Gamo:1997}). In this case one can expect that the sublattice symmetry is broken if the interaction strength between graphene and metal is strongly varied along the graphene lattice. However, as the bonding energy between graphene and metal is quite small (in the range of $50$\,-\,$150$\,meV/C-atom) and graphene is bonded to the metallic surface only via van der Waals (vdW) forces, the difference in the adsorption position cannot lead to the strong variation of the interaction strength between carbon atoms and the metallic substrate. Therefore the mechanism of the gap opening for the $\pi$ band around the Dirac point due to the violation of the sublattice symmetry in the graphene unit cell can be fully ruled out. 

\begin{figure}
\includegraphics[scale=2]{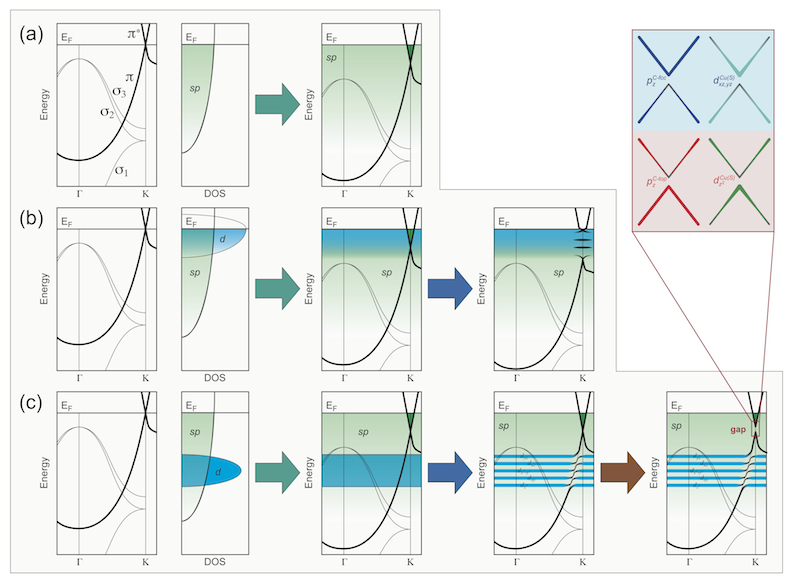}\\
\caption{\label{bands_scheme} Energy schemes demonstrating the discussed model: (a) graphene/$sp$-metal, (b) graphene/open-$d$-shell-metal, (c) graphene/closed-$d$-shell-metal. For detailed description, see text.} 
\end{figure}

First, let us consider a ``trivial'' case when graphene is placed on $sp$-metal [Fig.~\ref{bands_scheme}(a)]. Here the mobile $sp$ electrons fill the unoccupied $\pi^*$ states of graphene yielding the shift of $E_D$ below $E_F$ ($n$-doped graphene). The equilibrium distance as well as the doping level of graphene depends on the difference of the work functions of graphene and metallic surface~\cite{Giovannetti:2008,Khomyakov:2009}. The localization of the electron density on the graphene-derived $\pi$-orbitals at the graphene/metal interface leads to the increase of the attraction between graphene and metal. In the electronic structure of graphene the energy shift of $\pi$ and $\sigma$ states is equal. This description is valid for the graphene/alkali-metals~\cite{Petrovic:2013vz} and graphene/Al~\cite{Voloshina:2011NJP} interfaces, where simple $n$-doping of graphene is observed without any modification of the Dirac cone. In case of $p$-doped graphene (the electrons are transferred from graphene to metal), the Dirac cone is shifted upwards. Opposite to the former situation, the electron density depletion at the graphene/metal interface leads to the decrease of the vdW interaction at the interface as the dipole moment becomes smaller. Though $p$-doping is hardly possible when dealing with simple $sp$-metals, its formal reference here is needed for the further discussion.

If the underlying metal has an open $d$-shell, its interaction with graphene is more complicated [Fig.~\ref{bands_scheme}(b)]. At first, analogously to the case of $sp$-metals, the mobile $sp$ electrons of metal will define the initial doping of a graphene layer. In all existing cases doping of graphene leads to the positioning of $E_D$ within the $d$ band. Due to the existence of the $d$-orbitals with out-of-plane components ($d_{xz}$, $d_{yz}$, $d_{z^2}$), several, so-called, hybrid states are formed around the $\mathrm{K}$-point [Fig.~\ref{bands_scheme}(b)] from the corresponding $d$-states of metal and graphene-derived $\pi$-states. Here, several conditions for the hybridization are fulfilled, energy-, real-space-, and $k$-space-overlapping of the initial orbitals.

The effect of the $\pi$-$d$ hybridization leads to the violation of the sublattice symmetry in the graphene unit cell. Considering a graphene layer on top of the close-packed (111) surface of metal in the $top$-$fcc$ arrangement [Fig.~\ref{structure_orbitals}(b)]~\cite{Voloshina:2012c}, one can see that the $p_z$ orbital of the C-${top}$ atom overlaps with $d_{z^2}$ orbital of the metal atom in the surface layer, M(S), forming the $p_z^{C-top}$-\,$d_{z^2}$-hybrid. Similarly, the $p_z$ orbital of C-${fcc}$ atom overlaps with $d_{xz,yz}$ orbitals of M(S), forming the $p_z^{C-fcc}$-\,$d_{xz,yz}$-hybrid. Generally, for the free-standing graphene the electronic states from both carbon atoms are degenerate in the vicinity of the $\mathrm{K}$ point. In case of graphene adsorbed on metal, such degeneracy is lifted up via gap opening at $\mathrm{K}$ because the effect of hybridization described earlier leads to the formation of two hybrid states which cannot exist simultaneously at the same energy at the $\mathrm{K}$ point.

All reasonings given above for graphene on the $sp$ and open $d$-shell metals can be generalized for graphene on the closed $d$-shell metal [Fig.~\ref{bands_scheme}(c)]. Initial doping of graphene is governed by the mobile $sp$ electrons of metal. If $d$ band of metal is filled then it is located far below $E_D$ of the doped graphene. However, as was discussed earlier, the energy-, real-space-, and $k$-space-overlap of the metal $d$- and graphene-derived $\pi$-orbitals leads to the formation of the hybrid states and opening of the energy gaps according to the avoided-crossing mechanism in the energy and $k$-vector ranges where $\pi$ band is crossed by the $d$ bands. Such interaction, accompanied by the formation of $p_z^{C-top}$-\,$d_{z^2}$ and $p_z^{C-fcc}$-\,$d_{xz,yz}$ hybrid states, leads to the violation of the symmetry of the electronic states in the vicinity of the $\mathrm{K}$ point and opens the energy gap at $E_D$ of graphene. 

Thus, the effect of hybridization between graphene-derived $\pi$-states and filled shell $d$ states of metal, which exists in the energy range and for $k$-vector values far off the Dirac point, leads to the opening of the energy gap directly at $E_D$ due to the lifting of the degeneracy of the electronic states for two carbon sublattices: the $\pi$ states of carbon atoms from different sublattices hybridize with $d$ states of the different symmetries of the same interface metal atom, M(S) [Figs.~\ref{structure_orbitals}(b) and \ref{bands_scheme}(c)].

\begin{figure}
\includegraphics[scale=2.5]{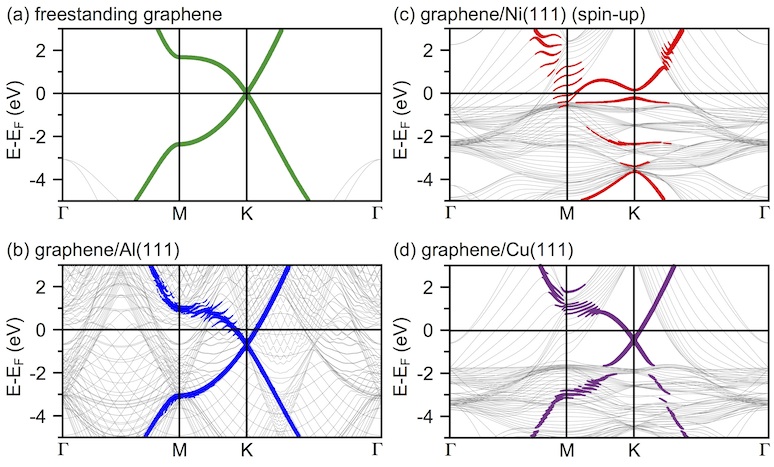}\\
\caption{\label{grgrNiAlCu} DFT calculated electronic structures of (a) free-standing graphene, (b) graphene/Al(111), (c) graphene/Ni(111), and (d) graphene/Cu(111). The weight of the C\,$p_z$ states in the band structures is proportional to the width of the colored line.} 
\end{figure}

Now we demonstrate the validity of our model with several representative examples. DFT calculations (see Supplementary material~\cite{suppl:all} for details) for free-standing graphene give a linear dispersion of the $\pi$ states in the vicinity of $E_F$ around the $\mathrm{K}$ point [Fig.~\ref{grgrNiAlCu}(a)]. Adsorption of graphene on Al(111) ($sp$ metal) yields the $n$-doping of graphene via transfer of the mobile $3s^2p^1$ electrons on the unoccupied $\pi^*$ graphene states and $E_D$ is placed at $E-E_F=-0.7$\,eV [Fig.~\ref{grgrNiAlCu}(b)]. This situation is described by the scheme shown in Fig.~\ref{bands_scheme}(a). The obtained results are in very good agreement with experimental data for the graphene/Al/Ni(111) system~\cite{Voloshina:2011NJP}, where simple doping of graphene without gap formation was observed. The similar results were experimentally found for other graphene/$sp$-metal systems~\cite{Gruneis:2008,Bianchi:2010cu,Petrovic:2013vz,Fedorov:2014aa}.

The electronic structure of graphene on Ni(111) (open $d$ shell metal) is shown in Fig.~\ref{grgrNiAlCu}(c). Here doping of graphene by $4s$ electrons places $E_D$ below $E_F$ that it is energetically overlaps with the Ni\,$3d$ bands. Electrostatic interaction decreases the distance between graphene and Ni that increases the space overlap of the C\,$p_z$ and Ni\,$3d$ orbitals. As a result of the energy-, real-space-, and $k$-space-overlap of the valence band states of graphene and metal, the Dirac cone of graphene is fully destroyed and several hybrid states are formed. Effect of hybridization decreases further the distance between graphene and Ni(111) to the equilibrium value of $2.08$\,\AA. This situation is described by the scheme in Fig.~\ref{bands_scheme}(b). These results are in very good agreement with experimental data for graphene/Ni(111) as well as for other ``strongly'' interacting graphene/metal systems~\cite{Gruneis:2008,Brugger:2009,Dedkov:2010a,Voloshina:2012c,Pacile:2013jc}.

The most interesting situation is observed for graphene/Cu(111) [Fig.~\ref{grgrNiAlCu}(d)], where graphene is $n$-doped due to the $s$ electrons transfer from Cu, and $E_D$ is at $E-E_F=-0.45$\,eV. Similar to graphene/Ni(111), doping of graphene decreases the distance between graphene and Cu(111). This effect allows to satisfy three necessary conditions for the hybridization of the Cu\,$3d$ and graphene\,$\pi$ states (space-, energy-, and $k$-vector conservation during hybridization) and the hybridization in this system is detected in the energy range of $E-E_F\approx-2 \ldots 4.5$\,eV [Fig.~\ref{grgrNiAlCu}(d)].

\begin{figure}
\includegraphics[scale=2.5]{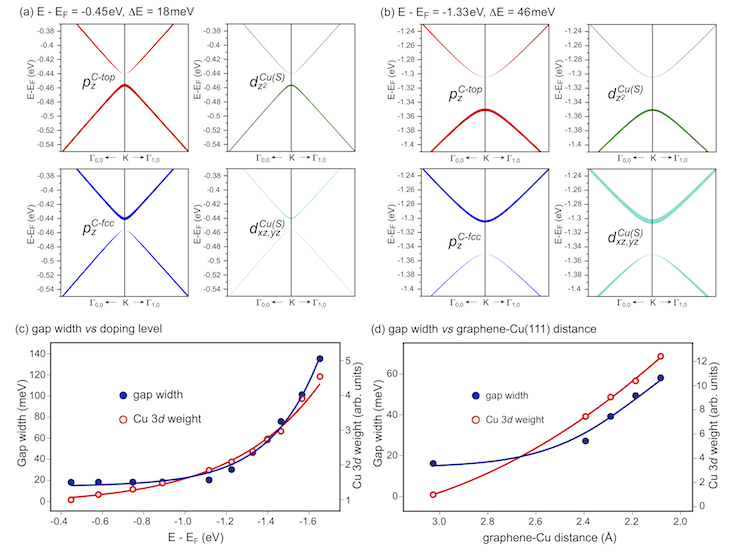}\\
\caption{\label{grCu_analysis} (a,b) Decomposition of the electronic structure of graphene/Cu(111) in the vicinity of the energy gap at the $\mathrm{K}$ point. The weight of the corresponding states is proportional to the width of the coloured line. (c) and (d) Dependencies of the gap width on the doping level of graphene and on the graphene-Cu(111) distance, respectively. Relative Cu\,$3d$ weight is plotted on the corresponding panels. The solid lines are eye-guides.} 
\end{figure}

The further analysis of the calculated band structure of graphene/Cu(111) shows that the energy gap of $18$\,meV is opened directly at $E_D$ [Fig.~\ref{grCu_analysis}(a)]. Decomposition of the bands around $E_D$ shows that the obtained picture is fully identical to the scheme presented in Fig.~\ref{bands_scheme}(c). The hybridization discussed earlier leads to the appearance of the hybrid states formed by the $p_z$ states of the two carbon atoms from different sublattices of the graphene layer and $3d$ states of different symmetry of the same top Cu atom: $p_z^{C-top}$-\,$d_{z^2}^{Cu(S)}$-hybrid and $p_z^{C-fcc}$-\,$d_{xz,yz}^{Cu(S)}$-hybrid. Formation of these two states leads to the lifting of the initial degeneracy at $E_D$, characteristic for the free-standing or doped graphene, and opening of the symmetry band gap.

Following the above discussion, one can expect that the width of the gap will further depend on the relative energy positions of the unperturbated Dirac cone and the metal $d$-band due to the increased partial weight of the metal $d$ states in the energy band: the closer the cone to the $d$-band the larger the gap at the Dirac point. Similar effect is expected if the distance between a graphene layer and Cu(111) is decreased due to increase of the space overlap of $p_z$ and $d$ orbitals.

In order to follow this effect we simulated the artificial doping of the graphene/Cu(111) system via adsorbing the Li atoms above the graphene layer. As expected the energy gap is increased and the results for the doping level of $E-E_F=-1.33$\,eV and the energy gap of $46$\,meV are presented in Fig.~\ref{grCu_analysis}(b), respectively. Further increase of the doping level leads to the widening of the energy gap as shown in the plot in Fig.~\ref{grCu_analysis}(c), where the relative Cu $3d$ weight in the lower-energy branch at the $\mathrm{K}$ point is also presented (for the corresponding band structures, see the Supplementary material, Fig.\,S2~\cite{suppl:all}). The similar effect is also observed if the distance between graphene and Cu(111) is varied: lessening the distance leads to the stronger space overlap of graphene\,$\pi$ and Cu\,$3d$ orbitals increasing the partial $d$ weight of the band and consequently further widening the energy gap for the $\pi$ states at the $\mathrm{K}$ point [Fig.~\ref{grCu_analysis}(d)] (for the corresponding band structures, see the Supplementary material, Fig.\,S3 and S4~\cite{suppl:all}). The calculated doping level for all distances from Fig.~\ref{grCu_analysis}(d) is below than the maximum doping obtained in Fig.~\ref{grCu_analysis}(c) as it is defined by the more mobile $4s$ electrons. From these data we can conclude that the energy overlap of the graphene\,$\pi$ and Cu\,$3d$ states plays a dominant role on the width of the band gap around $E_D$ of graphene.

It is interesting to note that appearance of the energy gap for the graphene-derived $\pi$ states and its width is caused by the mixing of the graphene and metal valence band states and it changes the band dispersion close to the border of the Brillouin zone drastically. As can be seen from Fig.~\ref{grCu_analysis} this leads to the violation of the linear dispersion of the $\pi$ states around the $\mathrm{K}$ point with the increase of the effective mass of carriers. This effect can drastically change the transport properties of the graphene-based devices where graphene/metal interfaces are implemented. 

The presented model describes the appearance of the energy gap at $E_D$ of graphene on metal on the qualitative level and predicts its behaviour as a function of the doping level of graphene in this system, that correlates with available experimental data. The doping of graphene alone cannot cause any sizeable gap in the electronic structure of graphene around $E_D$.

The three different cases considered in the present work are related to the $n$-doped graphene on metal. The situation is slightly different for the $p$-doped graphene on metal. Here the electron-transfer from graphene to metal reduces the polarization of the graphene layer that might lead to the reduction of the vdW attraction in the system that reduces the possible space overlapping of the graphene $p_z$ and metal $d$ orbitals. However, as shown in the experiment~\cite{Varykhalov:2010a,Enderlein:2010}, the doping of these systems can shift $E_D$ closer to the $d$ states of metal that widening the energy gap in the electronic structure of graphene.

\textit{In conclusion}, we have built an universal model for the description of the electronic structure of graphene on any metallic surface. Our model takes into account initial doping of a graphene layer by mobile $sp$ electrons, which depends on the work functions of graphene and metallic surface. Further modification of the electronic structure of graphene on $d$ metals is determined by the position of the Dirac point with respect of the $d$-band of metal. Such interaction leads either, in the case of the open $d$ shell metals, to the complete destroying of the Dirac cone around the $\mathrm{K}$ point or, in the case of the closed $d$ shell metals, to the opening the energy gap in the electronic structure of graphene directly at the Dirac point. Both effects are connected with the overlap of the graphene\,$p_z$ orbitals of two carbon atoms in the graphene unit cell with $d$ orbitals of the interface metal atom of different symmetries. Our funding explains all observed effects on the qualitative level and allows to predict the behaviour of the electronic states of graphene upon formation of the graphene/metal contacts in future devices. 

\bibliographystyle{apsrev}

\begin{thebibliography}{26}
\expandafter\ifx\csname natexlab\endcsname\relax\def\natexlab#1{#1}\fi
\expandafter\ifx\csname bibnamefont\endcsname\relax
  \def\bibnamefont#1{#1}\fi
\expandafter\ifx\csname bibfnamefont\endcsname\relax
  \def\bibfnamefont#1{#1}\fi
\expandafter\ifx\csname citenamefont\endcsname\relax
  \def\citenamefont#1{#1}\fi
\expandafter\ifx\csname url\endcsname\relax
  \def\url#1{\texttt{#1}}\fi
\expandafter\ifx\csname urlprefix\endcsname\relax\def\urlprefix{URL }\fi
\providecommand{\bibinfo}[2]{#2}
\providecommand{\eprint}[2][]{\url{#2}}

\bibitem[{\citenamefont{Novoselov et~al.}(2005)\citenamefont{Novoselov, Geim,
  Morozov, Jiang, Katsnelson, Grigorieva, Dubonos, and
  Firsov}}]{Novoselov:2005}
\bibinfo{author}{\bibfnamefont{K.}~\bibnamefont{Novoselov}},
  \bibinfo{author}{\bibfnamefont{A.}~\bibnamefont{Geim}},
  \bibinfo{author}{\bibfnamefont{S.}~\bibnamefont{Morozov}},
  \bibinfo{author}{\bibfnamefont{D.}~\bibnamefont{Jiang}},
  \bibinfo{author}{\bibfnamefont{M.}~\bibnamefont{Katsnelson}},
  \bibinfo{author}{\bibfnamefont{I.}~\bibnamefont{Grigorieva}},
  \bibinfo{author}{\bibfnamefont{S.}~\bibnamefont{Dubonos}}, \bibnamefont{and}
  \bibinfo{author}{\bibfnamefont{A.}~\bibnamefont{Firsov}},
  \bibinfo{journal}{Nature} \textbf{\bibinfo{volume}{438}},
  \bibinfo{pages}{197} (\bibinfo{year}{2005}).

\bibitem[{\citenamefont{Zhang et~al.}(2005)\citenamefont{Zhang, Tan, Stormer,
  and Kim}}]{Zhang:2005}
\bibinfo{author}{\bibfnamefont{Y.}~\bibnamefont{Zhang}},
  \bibinfo{author}{\bibfnamefont{Y.}~\bibnamefont{Tan}},
  \bibinfo{author}{\bibfnamefont{H.}~\bibnamefont{Stormer}}, \bibnamefont{and}
  \bibinfo{author}{\bibfnamefont{P.}~\bibnamefont{Kim}},
  \bibinfo{journal}{Nature} \textbf{\bibinfo{volume}{438}},
  \bibinfo{pages}{201} (\bibinfo{year}{2005}).

\bibitem[{\citenamefont{Novoselov et~al.}(2013)\citenamefont{Novoselov, ko,
  Colombo, Gellert, Schwab, and Kim}}]{Novoselov:2013hw}
\bibinfo{author}{\bibfnamefont{K.~S.} \bibnamefont{Novoselov}},
  \bibinfo{author}{\bibfnamefont{V.~I.} \bibnamefont{Fal'ko}},
  \bibinfo{author}{\bibfnamefont{L.}~\bibnamefont{Colombo}},
  \bibinfo{author}{\bibfnamefont{P.~R.} \bibnamefont{Gellert}},
  \bibinfo{author}{\bibfnamefont{M.~G.} \bibnamefont{Schwab}},
  \bibnamefont{and} \bibinfo{author}{\bibfnamefont{K.}~\bibnamefont{Kim}},
  \bibinfo{journal}{Nature} \textbf{\bibinfo{volume}{490}},
  \bibinfo{pages}{192} (\bibinfo{year}{2013}).

\bibitem[{\citenamefont{Bertoni et~al.}(2004)\citenamefont{Bertoni, Calmels,
  Altibelli, and Serin}}]{Bertoni:2004}
\bibinfo{author}{\bibfnamefont{G.}~\bibnamefont{Bertoni}},
  \bibinfo{author}{\bibfnamefont{L.}~\bibnamefont{Calmels}},
  \bibinfo{author}{\bibfnamefont{A.}~\bibnamefont{Altibelli}},
  \bibnamefont{and} \bibinfo{author}{\bibfnamefont{V.}~\bibnamefont{Serin}},
  \bibinfo{journal}{Phys. Rev. B} \textbf{\bibinfo{volume}{71}},
  \bibinfo{pages}{075402} (\bibinfo{year}{2004}).

\bibitem[{\citenamefont{Karpan et~al.}(2008)\citenamefont{Karpan, Khomyakov,
  Starikov, Giovannetti, Zwierzycki, Talanana, Brocks, Brink, and
  Kelly}}]{Karpan:2008}
\bibinfo{author}{\bibfnamefont{V.~M.} \bibnamefont{Karpan}},
  \bibinfo{author}{\bibfnamefont{P.~A.} \bibnamefont{Khomyakov}},
  \bibinfo{author}{\bibfnamefont{A.~A.} \bibnamefont{Starikov}},
  \bibinfo{author}{\bibfnamefont{G.}~\bibnamefont{Giovannetti}},
  \bibinfo{author}{\bibfnamefont{M.}~\bibnamefont{Zwierzycki}},
  \bibinfo{author}{\bibfnamefont{M.}~\bibnamefont{Talanana}},
  \bibinfo{author}{\bibfnamefont{G.}~\bibnamefont{Brocks}},
  \bibinfo{author}{\bibfnamefont{J.~v.~d.} \bibnamefont{Brink}},
  \bibnamefont{and} \bibinfo{author}{\bibfnamefont{P.~J.} \bibnamefont{Kelly}},
  \bibinfo{journal}{Phys. Rev. B} \textbf{\bibinfo{volume}{78}},
  \bibinfo{pages}{195419} (\bibinfo{year}{2008}).

\bibitem[{\citenamefont{Weser et~al.}(2011)\citenamefont{Weser, Voloshina,
  Horn, and Dedkov}}]{Weser:2011}
\bibinfo{author}{\bibfnamefont{M.}~\bibnamefont{Weser}},
  \bibinfo{author}{\bibfnamefont{E.~N.} \bibnamefont{Voloshina}},
  \bibinfo{author}{\bibfnamefont{K.}~\bibnamefont{Horn}}, \bibnamefont{and}
  \bibinfo{author}{\bibfnamefont{Y.~S.} \bibnamefont{Dedkov}},
  \bibinfo{journal}{Phys. Chem. Chem. Phys.} \textbf{\bibinfo{volume}{13}},
  \bibinfo{pages}{7534} (\bibinfo{year}{2011}).

\bibitem[{\citenamefont{Dedkov and Fonin}(2010)}]{Dedkov:2010a}
\bibinfo{author}{\bibfnamefont{Y.~S.} \bibnamefont{Dedkov}} \bibnamefont{and}
  \bibinfo{author}{\bibfnamefont{M.}~\bibnamefont{Fonin}},
  \bibinfo{journal}{New J. Phys.} \textbf{\bibinfo{volume}{12}},
  \bibinfo{pages}{125004} (\bibinfo{year}{2010}).

\bibitem[{\citenamefont{Varykhalov and Rader}(2009)}]{Varykhalov:2009}
\bibinfo{author}{\bibfnamefont{A.}~\bibnamefont{Varykhalov}} \bibnamefont{and}
  \bibinfo{author}{\bibfnamefont{O.}~\bibnamefont{Rader}},
  \bibinfo{journal}{Phys. Rev. B} \textbf{\bibinfo{volume}{80}},
  \bibinfo{pages}{035437} (\bibinfo{year}{2009}).

\bibitem[{\citenamefont{Brugger et~al.}(2009)\citenamefont{Brugger, Guenther,
  Wang, Dil, Bocquet, Osterwalder, Wintterlin, and Greber}}]{Brugger:2009}
\bibinfo{author}{\bibfnamefont{T.}~\bibnamefont{Brugger}},
  \bibinfo{author}{\bibfnamefont{S.}~\bibnamefont{Guenther}},
  \bibinfo{author}{\bibfnamefont{B.}~\bibnamefont{Wang}},
  \bibinfo{author}{\bibfnamefont{J.~H.} \bibnamefont{Dil}},
  \bibinfo{author}{\bibfnamefont{M.-L.} \bibnamefont{Bocquet}},
  \bibinfo{author}{\bibfnamefont{J.}~\bibnamefont{Osterwalder}},
  \bibinfo{author}{\bibfnamefont{J.}~\bibnamefont{Wintterlin}},
  \bibnamefont{and} \bibinfo{author}{\bibfnamefont{T.}~\bibnamefont{Greber}},
  \bibinfo{journal}{Phys. Rev. B} \textbf{\bibinfo{volume}{79}},
  \bibinfo{pages}{045407} (\bibinfo{year}{2009}).

\bibitem[{\citenamefont{Dedkov et~al.}(2001)\citenamefont{Dedkov, Shikin,
  Adamchuk, Molodtsov, Laubschat, Bauer, and Kaindl}}]{Dedkov:2001}
\bibinfo{author}{\bibfnamefont{Y.~S.} \bibnamefont{Dedkov}},
  \bibinfo{author}{\bibfnamefont{A.~M.} \bibnamefont{Shikin}},
  \bibinfo{author}{\bibfnamefont{V.~K.} \bibnamefont{Adamchuk}},
  \bibinfo{author}{\bibfnamefont{S.~L.} \bibnamefont{Molodtsov}},
  \bibinfo{author}{\bibfnamefont{C.}~\bibnamefont{Laubschat}},
  \bibinfo{author}{\bibfnamefont{A.}~\bibnamefont{Bauer}}, \bibnamefont{and}
  \bibinfo{author}{\bibfnamefont{G.}~\bibnamefont{Kaindl}},
  \bibinfo{journal}{Phys. Rev. B} \textbf{\bibinfo{volume}{64}},
  \bibinfo{pages}{035405} (\bibinfo{year}{2001}).

\bibitem[{\citenamefont{Pletikosi{\'c}
  et~al.}(2009)\citenamefont{Pletikosi{\'c}, Kralj, Pervan, Brako, Coraux,
  N'diaye, Busse, and Michely}}]{Pletikosic:2009}
\bibinfo{author}{\bibfnamefont{I.}~\bibnamefont{Pletikosi{\'c}}},
  \bibinfo{author}{\bibfnamefont{M.}~\bibnamefont{Kralj}},
  \bibinfo{author}{\bibfnamefont{P.}~\bibnamefont{Pervan}},
  \bibinfo{author}{\bibfnamefont{R.}~\bibnamefont{Brako}},
  \bibinfo{author}{\bibfnamefont{J.}~\bibnamefont{Coraux}},
  \bibinfo{author}{\bibfnamefont{A.}~\bibnamefont{N'diaye}},
  \bibinfo{author}{\bibfnamefont{C.}~\bibnamefont{Busse}}, \bibnamefont{and}
  \bibinfo{author}{\bibfnamefont{T.}~\bibnamefont{Michely}},
  \bibinfo{journal}{Phys. Rev. Lett.} \textbf{\bibinfo{volume}{102}},
  \bibinfo{pages}{056808} (\bibinfo{year}{2009}).

\bibitem[{\citenamefont{Sutter et~al.}(2009)\citenamefont{Sutter, Sadowski, and
  Sutter}}]{Sutter:2009a}
\bibinfo{author}{\bibfnamefont{P.}~\bibnamefont{Sutter}},
  \bibinfo{author}{\bibfnamefont{J.}~\bibnamefont{Sadowski}}, \bibnamefont{and}
  \bibinfo{author}{\bibfnamefont{E.}~\bibnamefont{Sutter}},
  \bibinfo{journal}{Phys. Rev. B} \textbf{\bibinfo{volume}{80}},
  \bibinfo{pages}{245411} (\bibinfo{year}{2009}).

\bibitem[{\citenamefont{Varykhalov et~al.}(2010)\citenamefont{Varykhalov,
  Scholz, Kim, and Rader}}]{Varykhalov:2010a}
\bibinfo{author}{\bibfnamefont{A.}~\bibnamefont{Varykhalov}},
  \bibinfo{author}{\bibfnamefont{M.}~\bibnamefont{Scholz}},
  \bibinfo{author}{\bibfnamefont{T.}~\bibnamefont{Kim}}, \bibnamefont{and}
  \bibinfo{author}{\bibfnamefont{O.}~\bibnamefont{Rader}},
  \bibinfo{journal}{Phys. Rev. B} \textbf{\bibinfo{volume}{82}},
  \bibinfo{pages}{121101} (\bibinfo{year}{2010}).

\bibitem[{\citenamefont{Voloshina et~al.}(2011)\citenamefont{Voloshina,
  Generalov, Weser, B{\"o}ttcher, Horn, and Dedkov}}]{Voloshina:2011NJP}
\bibinfo{author}{\bibfnamefont{E.~N.} \bibnamefont{Voloshina}},
  \bibinfo{author}{\bibfnamefont{A.}~\bibnamefont{Generalov}},
  \bibinfo{author}{\bibfnamefont{M.}~\bibnamefont{Weser}},
  \bibinfo{author}{\bibfnamefont{S.}~\bibnamefont{B{\"o}ttcher}},
  \bibinfo{author}{\bibfnamefont{K.}~\bibnamefont{Horn}}, \bibnamefont{and}
  \bibinfo{author}{\bibfnamefont{Y.~S.} \bibnamefont{Dedkov}},
  \bibinfo{journal}{New J. Phys.} \textbf{\bibinfo{volume}{13}},
  \bibinfo{pages}{113028} (\bibinfo{year}{2011}).

\bibitem[{\citenamefont{Voloshina and Dedkov}(2012)}]{Voloshina:2012c}
\bibinfo{author}{\bibfnamefont{E.}~\bibnamefont{Voloshina}} \bibnamefont{and}
  \bibinfo{author}{\bibfnamefont{Y.}~\bibnamefont{Dedkov}},
  \bibinfo{journal}{Phys. Chem. Chem. Phys.} \textbf{\bibinfo{volume}{14}},
  \bibinfo{pages}{13502} (\bibinfo{year}{2012}).

\bibitem[{\citenamefont{Shikin et~al.}(2013)\citenamefont{Shikin, Rybkin,
  Marchenko, Rybkina, Scholz, Rader, and Varykhalov}}]{Shikin:2013fr}
\bibinfo{author}{\bibfnamefont{A.~M.} \bibnamefont{Shikin}},
  \bibinfo{author}{\bibfnamefont{A.~G.} \bibnamefont{Rybkin}},
  \bibinfo{author}{\bibfnamefont{D.}~\bibnamefont{Marchenko}},
  \bibinfo{author}{\bibfnamefont{A.~A.} \bibnamefont{Rybkina}},
  \bibinfo{author}{\bibfnamefont{M.~R.} \bibnamefont{Scholz}},
  \bibinfo{author}{\bibfnamefont{O.}~\bibnamefont{Rader}}, \bibnamefont{and}
  \bibinfo{author}{\bibfnamefont{A.}~\bibnamefont{Varykhalov}},
  \bibinfo{journal}{New J. Phys.} \textbf{\bibinfo{volume}{15}},
  \bibinfo{pages}{013016} (\bibinfo{year}{2013}).

\bibitem[{\citenamefont{Gr{\"u}neis and Vyalikh}(2008)}]{Gruneis:2008}
\bibinfo{author}{\bibfnamefont{A.}~\bibnamefont{Gr{\"u}neis}} \bibnamefont{and}
  \bibinfo{author}{\bibfnamefont{D.}~\bibnamefont{Vyalikh}},
  \bibinfo{journal}{Phys. Rev. B} \textbf{\bibinfo{volume}{77}},
  \bibinfo{pages}{193401} (\bibinfo{year}{2008}).

\bibitem[{\citenamefont{Petrovi{\'c} et~al.}(2013)\citenamefont{Petrovi{\'c},
  Raki{\'c}, Runte, Busse, Sadowski, Lazic, Pletikosi{\'c}, Pan, Milun, and
  Pervan}}]{Petrovic:2013vz}
\bibinfo{author}{\bibfnamefont{M.}~\bibnamefont{Petrovi{\'c}}},
  \bibinfo{author}{\bibfnamefont{I.~{\v S}.} \bibnamefont{Raki{\'c}}},
  \bibinfo{author}{\bibfnamefont{S.}~\bibnamefont{Runte}},
  \bibinfo{author}{\bibfnamefont{C.}~\bibnamefont{Busse}},
  \bibinfo{author}{\bibfnamefont{J.~T.} \bibnamefont{Sadowski}},
  \bibinfo{author}{\bibfnamefont{P.}~\bibnamefont{Lazic}},
  \bibinfo{author}{\bibfnamefont{I.}~\bibnamefont{Pletikosi{\'c}}},
  \bibinfo{author}{\bibfnamefont{Z.~H.} \bibnamefont{Pan}},
  \bibinfo{author}{\bibfnamefont{M.}~\bibnamefont{Milun}}, \bibnamefont{and}
  \bibinfo{author}{\bibfnamefont{P.}~\bibnamefont{Pervan}},
  \bibinfo{journal}{Nature Communications} \textbf{\bibinfo{volume}{4}}
  (\bibinfo{year}{2013}).

\bibitem[{\citenamefont{Fedorov et~al.}(2014)\citenamefont{Fedorov, Verbitskiy,
  Haberer, Struzzi, Petaccia, Usachov, Vilkov, Vyalikh, Fink, Knupfer
  et~al.}}]{Fedorov:2014aa}
\bibinfo{author}{\bibfnamefont{A.~V.} \bibnamefont{Fedorov}},
  \bibinfo{author}{\bibfnamefont{N.~I.} \bibnamefont{Verbitskiy}},
  \bibinfo{author}{\bibfnamefont{D.}~\bibnamefont{Haberer}},
  \bibinfo{author}{\bibfnamefont{C.}~\bibnamefont{Struzzi}},
  \bibinfo{author}{\bibfnamefont{L.}~\bibnamefont{Petaccia}},
  \bibinfo{author}{\bibfnamefont{D.}~\bibnamefont{Usachov}},
  \bibinfo{author}{\bibfnamefont{O.~Y.} \bibnamefont{Vilkov}},
  \bibinfo{author}{\bibfnamefont{D.~V.} \bibnamefont{Vyalikh}},
  \bibinfo{author}{\bibfnamefont{J.}~\bibnamefont{Fink}},
  \bibinfo{author}{\bibfnamefont{M.}~\bibnamefont{Knupfer}},
  \bibnamefont{et~al.}, \bibinfo{journal}{Nature Communications}
  \textbf{\bibinfo{volume}{5}}, \bibinfo{pages}{3257} (\bibinfo{year}{2014}).

\bibitem[{sup()}]{suppl:all}
\bibinfo{journal}{Supplementary material}  (????).

\bibitem[{\citenamefont{Gamo et~al.}(1997)\citenamefont{Gamo, Nagashima,
  Wakabayashi, Terai, and Oshima}}]{Gamo:1997}
\bibinfo{author}{\bibfnamefont{Y.}~\bibnamefont{Gamo}},
  \bibinfo{author}{\bibfnamefont{A.}~\bibnamefont{Nagashima}},
  \bibinfo{author}{\bibfnamefont{M.}~\bibnamefont{Wakabayashi}},
  \bibinfo{author}{\bibfnamefont{M.}~\bibnamefont{Terai}}, \bibnamefont{and}
  \bibinfo{author}{\bibfnamefont{C.}~\bibnamefont{Oshima}},
  \bibinfo{journal}{Surf. Sci.} \textbf{\bibinfo{volume}{374}},
  \bibinfo{pages}{61} (\bibinfo{year}{1997}).

\bibitem[{\citenamefont{Giovannetti et~al.}(2008)\citenamefont{Giovannetti,
  Khomyakov, Brocks, Karpan, Brink, and Kelly}}]{Giovannetti:2008}
\bibinfo{author}{\bibfnamefont{G.}~\bibnamefont{Giovannetti}},
  \bibinfo{author}{\bibfnamefont{P.~A.} \bibnamefont{Khomyakov}},
  \bibinfo{author}{\bibfnamefont{G.}~\bibnamefont{Brocks}},
  \bibinfo{author}{\bibfnamefont{V.~M.} \bibnamefont{Karpan}},
  \bibinfo{author}{\bibfnamefont{J.~v.~d.} \bibnamefont{Brink}},
  \bibnamefont{and} \bibinfo{author}{\bibfnamefont{P.~J.} \bibnamefont{Kelly}},
  \bibinfo{journal}{Phys. Rev. Lett.} \textbf{\bibinfo{volume}{101}},
  \bibinfo{pages}{026803} (\bibinfo{year}{2008}).

\bibitem[{\citenamefont{Khomyakov et~al.}(2009)\citenamefont{Khomyakov,
  Giovannetti, Rusu, Brocks, Brink, and Kelly}}]{Khomyakov:2009}
\bibinfo{author}{\bibfnamefont{P.~A.} \bibnamefont{Khomyakov}},
  \bibinfo{author}{\bibfnamefont{G.}~\bibnamefont{Giovannetti}},
  \bibinfo{author}{\bibfnamefont{P.~C.} \bibnamefont{Rusu}},
  \bibinfo{author}{\bibfnamefont{G.}~\bibnamefont{Brocks}},
  \bibinfo{author}{\bibfnamefont{J.~v.~d.} \bibnamefont{Brink}},
  \bibnamefont{and} \bibinfo{author}{\bibfnamefont{P.~J.} \bibnamefont{Kelly}},
  \bibinfo{journal}{Phys. Rev. B} \textbf{\bibinfo{volume}{79}},
  \bibinfo{pages}{195425} (\bibinfo{year}{2009}).

\bibitem[{\citenamefont{Bianchi et~al.}(2010)\citenamefont{Bianchi, Rienks,
  Lizzit, Baraldi, Balog, Hornekaer, and Hofmann}}]{Bianchi:2010cu}
\bibinfo{author}{\bibfnamefont{M.}~\bibnamefont{Bianchi}},
  \bibinfo{author}{\bibfnamefont{E.~D.~L.} \bibnamefont{Rienks}},
  \bibinfo{author}{\bibfnamefont{S.}~\bibnamefont{Lizzit}},
  \bibinfo{author}{\bibfnamefont{A.}~\bibnamefont{Baraldi}},
  \bibinfo{author}{\bibfnamefont{R.}~\bibnamefont{Balog}},
  \bibinfo{author}{\bibfnamefont{L.}~\bibnamefont{Hornekaer}},
  \bibnamefont{and} \bibinfo{author}{\bibfnamefont{P.}~\bibnamefont{Hofmann}},
  \bibinfo{journal}{Phys. Rev. B} \textbf{\bibinfo{volume}{81}},
  \bibinfo{pages}{041403} (\bibinfo{year}{2010}).

\bibitem[{\citenamefont{Pacil{\'e} et~al.}(2013)\citenamefont{Pacil{\'e},
  Leicht, Papagno, Sheverdyaeva, Moras, Carbone, Krausert, Zielke, Fonin,
  Dedkov et~al.}}]{Pacile:2013jc}
\bibinfo{author}{\bibfnamefont{D.}~\bibnamefont{Pacil{\'e}}},
  \bibinfo{author}{\bibfnamefont{P.}~\bibnamefont{Leicht}},
  \bibinfo{author}{\bibfnamefont{M.}~\bibnamefont{Papagno}},
  \bibinfo{author}{\bibfnamefont{P.~M.} \bibnamefont{Sheverdyaeva}},
  \bibinfo{author}{\bibfnamefont{P.}~\bibnamefont{Moras}},
  \bibinfo{author}{\bibfnamefont{C.}~\bibnamefont{Carbone}},
  \bibinfo{author}{\bibfnamefont{K.}~\bibnamefont{Krausert}},
  \bibinfo{author}{\bibfnamefont{L.}~\bibnamefont{Zielke}},
  \bibinfo{author}{\bibfnamefont{M.}~\bibnamefont{Fonin}},
  \bibinfo{author}{\bibfnamefont{Y.~S.} \bibnamefont{Dedkov}},
  \bibnamefont{et~al.}, \bibinfo{journal}{Phys. Rev. B}
  \textbf{\bibinfo{volume}{87}}, \bibinfo{pages}{035420}
  (\bibinfo{year}{2013}).

\bibitem[{\citenamefont{Enderlein et~al.}(2010)\citenamefont{Enderlein, Kim,
  Bostwick, Rotenberg, and Horn}}]{Enderlein:2010}
\bibinfo{author}{\bibfnamefont{C.}~\bibnamefont{Enderlein}},
  \bibinfo{author}{\bibfnamefont{Y.~S.} \bibnamefont{Kim}},
  \bibinfo{author}{\bibfnamefont{A.}~\bibnamefont{Bostwick}},
  \bibinfo{author}{\bibfnamefont{E.}~\bibnamefont{Rotenberg}},
  \bibnamefont{and} \bibinfo{author}{\bibfnamefont{K.}~\bibnamefont{Horn}},
  \bibinfo{journal}{New J. Phys.} \textbf{\bibinfo{volume}{12}},
  \bibinfo{pages}{033014} (\bibinfo{year}{2010}).

\end{thebibliography}

\begin{thebibliography}{99}
\bibitem{paw}
P. E. Bl\"ochl, Phys. Rev. B \textbf{50}, 17953 (1994).
\bibitem{pbe}
J. P. Perdew, K. Burke, and M. Ernzerhof, Phys.Rev.Lett. \textbf{77}, 3865 (1996).
\bibitem{vasp}
G. Kresse and J. Hafner, J. Phys.: Condens. Matter \textbf{6}, 8245 (1994).
\bibitem{grimme}
S. Grimme, J. Comput. Chem. \textbf{27}, 1787 (2006).
\end{thebibliography}

\clearpage
\begin{figure}
\includegraphics[scale=3]{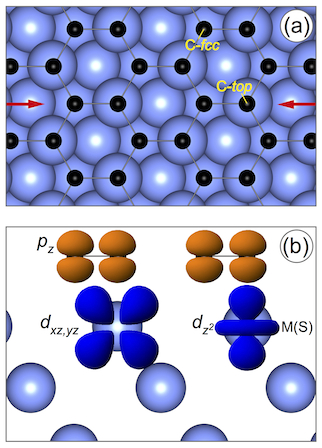}\\
\caption{\label{structure_orbitals} (a) Top and (b) side views of the graphene/M(111) interface (side view was taken as a cut through atoms marked by arrows). In (b) the C\,$p_z$ and metal $d_{z^2}$, $d_{xz}$ orbitals are shown for graphene and top metallic layer, respectively.} 
\end{figure}

\clearpage

\clearpage

\clearpage

\clearpage
\noindent
Supplementary material for manuscript:\\
\textbf{General approach to the understanding the electronic structure of graphene on metals}\\
\newline
E. N. Voloshina$^1$ and Yu. S. Dedkov$^2$\\
\newline
$^1$Institut f\"ur Chemie, Humboldt-Universit\"at zu Berlin, 10099 Berlin, Germany\\
$^2$SPECS Surface Nano Analysis GmbH, Voltastra\ss e 5, 13355 Berlin, Germany
\newline
\newline
\textbf{Content}
\begin{itemize}
\item \textbf{Comparison of the $(1\times1)$graphene/$(1\times1)$metal and $(n\times n)$graphene/$(m\times m)$metal structures.}

\item \textbf{Description of computational details.}

\item \textbf{Fig.\,S1.} Scheme for the formation of the energy gap around $E_D$ via lifting the degeneracy of the electronic states of graphene on closed $d$-shell metal.

\item \textbf{Fig.\,S2.} Band structures of graphene/Cu(111) for different doping levels. The corresponding energy gaps are marked in the figure. 

\item\textbf{Fig.\,S3.} Band structures of graphene/Cu(111) for different distances between graphene and Cu(111). The corresponding energy gaps are marked in the figure. 

\item\textbf{Figs.\,S4.} Analysis of the electronic structure of graphene/Cu(111) around the $\mathrm{K}$ point corresponding to the graphene-Cu(111) distance of $d=2.0828$\,\AA.
\end{itemize}

\clearpage
\noindent\textbf{Comparison of the $(1\times1)$graphene/$(1\times1)$metal and $(n\times n)$graphene/$(m\times m)$metal structures.}

\begin{figure}
\includegraphics[scale=2]{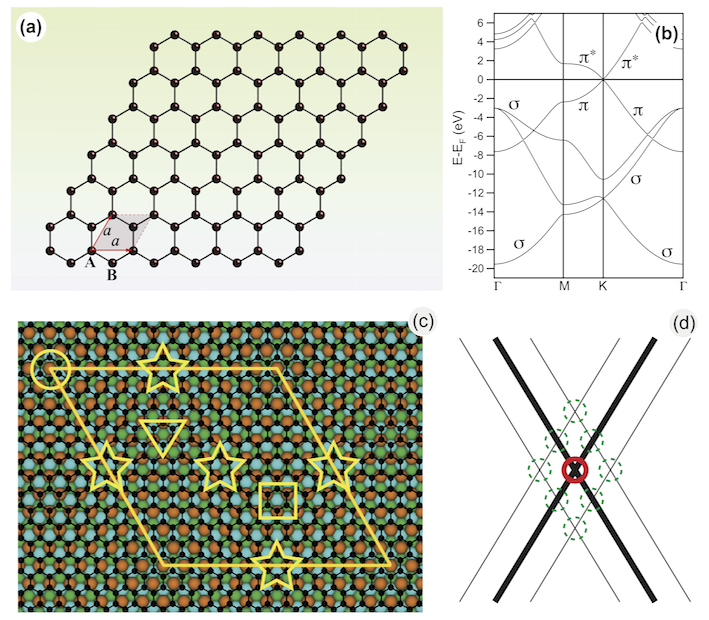}\\
\caption{\label{figSgraphene_1} (a) Crystallographic structure of free-standing graphene. (b) DFT computed band structure of free-standing graphene. (c) Crystallographic structure of the $(12\times12)$graphene on $(11\times11)$metal. (d) Main electronic bands of graphene corresponding to the Brillouin zone of graphene are shown by thick solid lines. Replica bands due to the moir\'e structure are shown by the thin lines. Position for the gap at the Dirac point of graphene is shown by the red circle. Places for the additional mini-gaps are shown by the green dashed circles.} 
\end{figure}

Graphene is a 2D layer of carbon atoms arranged in the honeycomb lattice (two carbon atoms per unit cell) [Fig.~\ref{figSgraphene_1}(a)]. The electronic states in free-standing graphene has a linear dispersion around the Dirac point at the Fermi level [Fig.~\ref{figSgraphene_1}(b)]. For free-standing graphene the states at the Dirac point degenerate as the two carbon atoms are fully identical and if there is a potential variation along the graphene unit cell, then the symmetry of states which belong to the sublattice A and B is different and the band gap is opened at the Dirac point. In this case the appearance of the band gap at the Dirac point of graphene is defined by the violation of the sublattice local symmetry in the graphene layer. This is one of the basic statements in the physics of graphene. 

If graphene is adsorbed on metal, then in reality the formation of the ideal $(1\times1)$ lattice-matched graphene-metal interface is unlikely. The closest to this situation are graphene/Ni(111) and graphene/Co(0001), which have a lattice difference of about 1-1.5\%. In most cases the so-called moir\'e structures are realised for the graphene-metal interfaces [Fig.~\ref{figSgraphene_1}(c)]. 

For free-standing graphene the effect of the moir\'e structure can lead to the formation of the additional replica bands [thin solid lines in Fig.~\ref{figSgraphene_1}(d)] in the electronic structure of graphene [main bands of graphene corresponding to the Brillouin zone of graphene are shown by the thick solid lines in Fig.~\ref{figSgraphene_1}(d)] and as a consequence to the formation of a mini-gaps due to avoided crossing mechanism [green dashed circles in Fig.~\ref{figSgraphene_1}(d)]. These moir\'e structures of any periodicity cannot influence (in any way!) the main mechanism of the gap formation at the Dirac point of graphene via violation of a sublattice symmetry in the graphene unit cell [red circle in Fig.~\ref{figSgraphene_1}(d)].

\begin{figure}
\includegraphics[scale=1.8]{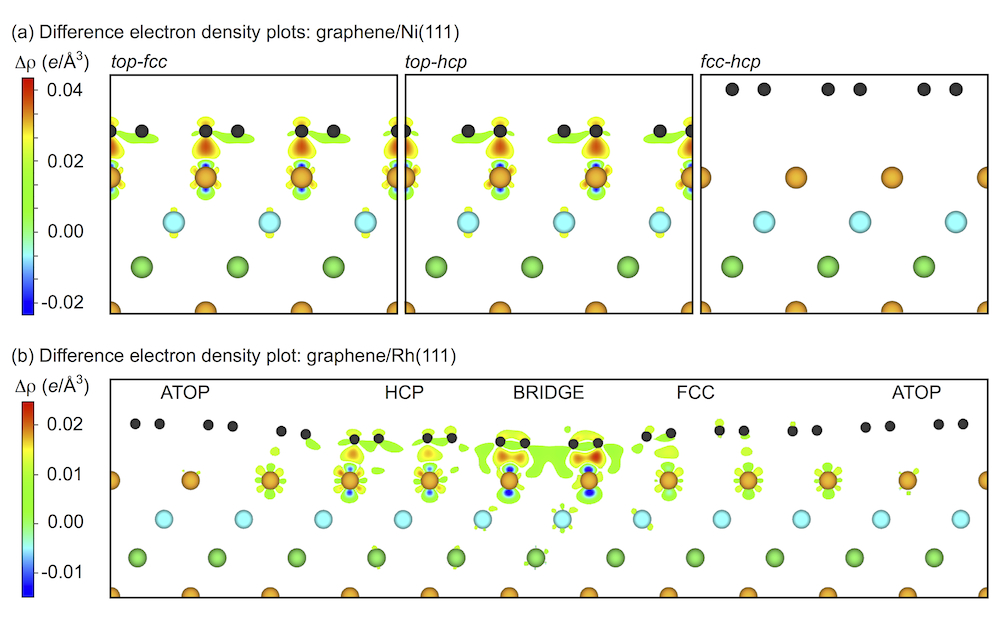}\\
\caption{\label{figSgraphene_2} (b) and (c) show the side views as well as the difference electron density, $\Delta \rho(r) = \rho_{gr/Rh} - \rho_{Rh} - \rho_{gr}$, plotted in units of $e/$\AA$^3$ calculated for different positions of graphene-Ni(111) and graphene-Rh(111). Red (blue) color indicates regions where the electron density increases (decreases). This figure is taken from E. Voloshina and Yu. Dedkov, Phys. Chem. Chem. Phys. \textbf{14}, 13502 (2012).} 
\end{figure}

In case of the graphene adsorption on metal it is obvious that any moir\'e graphene-metal system consists of the places where different high-symmetry stackings are realised [Fig.~\ref{figSgraphene_1}(c) and Fig.~\ref{figSgraphene_2}]. Fig.~\ref{figSgraphene_1}(c) shows the crystallographic structure of the $(12\times12)$graphene layer on top of $(11\times11)$ close packed metallic slab. One can see that this structure has 4 high-symmetry places, the so called ATOP, FCC, HCP, and BRIDGE as well as the intermediate positions. For the further description of this model we refer to recent manuscript [Phys. Chem. Chem. Phys. \textbf{14}, 13502 (2012)]. For the high-symmetry positions we have situations that they are very close to the $(1\times1)$ structures of the corresponding stacking. See for example Fig.~\ref{figSgraphene_2} where moir\'e structure of graphene/Rh(111) is compared with the corresponding high-symmetry stackings of graphene/Ni(111): ATOP $\rightarrow$ fcc-hcp, HCP $\rightarrow$ top-fcc, and FCC $\rightarrow$ top-hcp. It is obvious and was intensively discussed in the literature that the electronic structure of graphene in its complete Brillouin zone (corresponding to the lattice constant of 2.46\,\AA) is defined by the bonding strength and the electronic structure at the most perturbed graphene places. For example, in the case of graphene/Rh(111), by HCP, FCC, and BRIDGE positions. In all these places, which can be considered for simplicity as $(1\times1)$ the local symmetry in the graphene unit cell is broken as the $p_z$ orbitals of different carbon atoms overlap with $d$ states (which have $z$ component in the wave function) of the underlying top metal layer of the different symmetry. This effect as shown in the present manuscript leads to the opening of the energy gap at the Dirac point of graphene. The moir\'e structure is the second step in the gap formation which can lead to the decreasing or increasing of the energy gap (depending on the size of the structure as well as on the respective orientation angle), but can not be considered as a mechanism for the gap formation at the Dirac point. Here we would like to point again, that appearance of the energy gap at the Dirac point has a local nature due to the violation of the local sublattice symmetry in the graphene layer.

\clearpage
\noindent\textbf{Description of computational details.}

The DFT calculations were carried out using the projector augmented wave method~\cite{paw}, a plane wave basis set and the generalized gradient approximation as parameterized by Perdew \textit{et al.} (PBE)~\cite{pbe}, as implemented in the VASP program~\cite{vasp}. The plane wave kinetic energy cutoff was set to $500$\,eV. The long-range van der Waals interactions were accounted for by means of a semiempirical DFT-D2 approach~\cite{grimme}. In the total energy calculations and during the structural relaxation (the positions of the carbon atoms as well as those of the top two layers of metal are optimized) the $k$-meshes for sampling of the supercell Brillouin zone were chosen to be as dense as $24\times 24$ and $12\times 12$, respectively, when folded up to the simple graphene unit cell. The studied systems are modelled using a supercell consisting of 13 layers of metal atoms and a graphene sheet adsorbed on both sides of the slab. Metallic slab replicas are separated by about $24$\,\AA in the surface normal direction, leading to an effective vacuum region of at about $18$\,\AA. Graphene/Ni(111) and graphene/Cu(111) were modelled as a lattice-matched systems with $(1\times1)$ periodicity in a most stable $top-fcc$ configurations with a lattice constant corresponding to graphene. Graphene/Al(111) was modelled a lattice-matched system with $(2\times2)$ periodicity with respect to graphene lattice. In this structure the carbon atoms occupy all possible high-symmetry position of Al(111).

\clearpage
\begin{figure}
\includegraphics[scale=3]{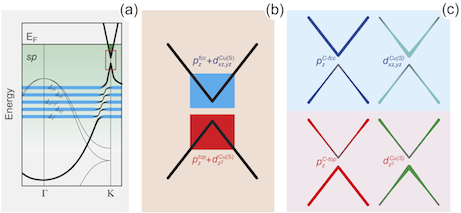}\\
\end{figure}
\noindent\textbf{Fig.\,S1.} Scheme for the formation of the energy gap around $E_D$ via lifting the degeneracy of the electronic states of graphene on closed $d$-shell metal.

\clearpage
\begin{figure}
\includegraphics[scale=2]{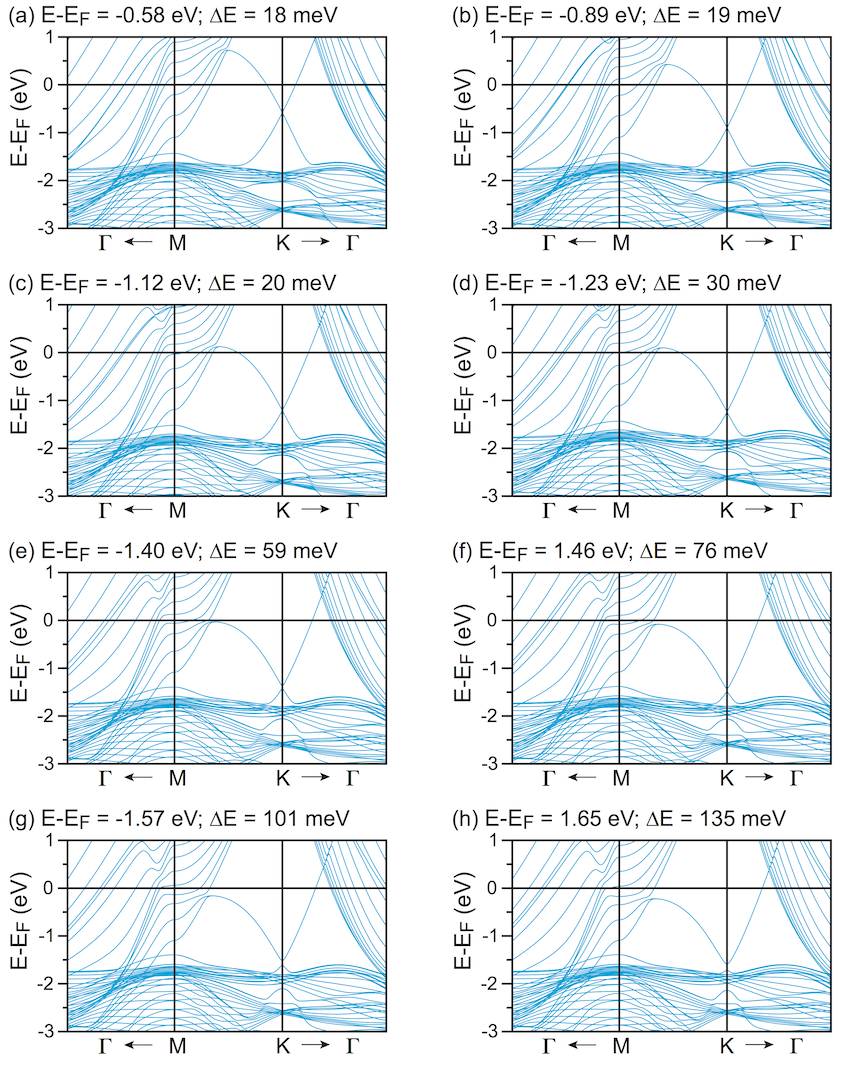}\\
\end{figure}
\noindent\textbf{Fig.\,S2.} Band structures of graphene/Cu(111) for different doping levels. The corresponding energy gaps are marked in the figure.  

\clearpage
\begin{figure}
\includegraphics[scale=2]{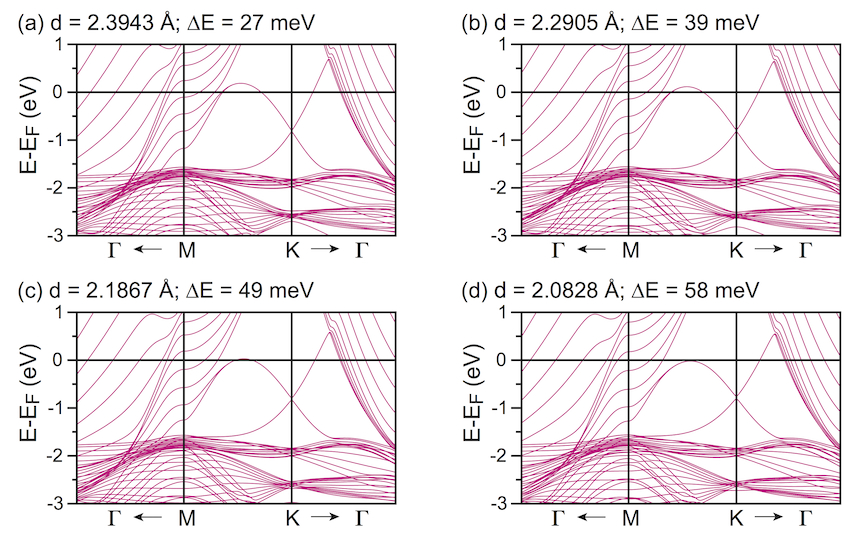}\\
\end{figure}
\noindent\textbf{Fig.\,S3.} Band structures of graphene/Cu(111) for different distances between graphene and Cu(111). The corresponding energy gaps are marked in the figure. 

\clearpage
\begin{figure}
\includegraphics[scale=2]{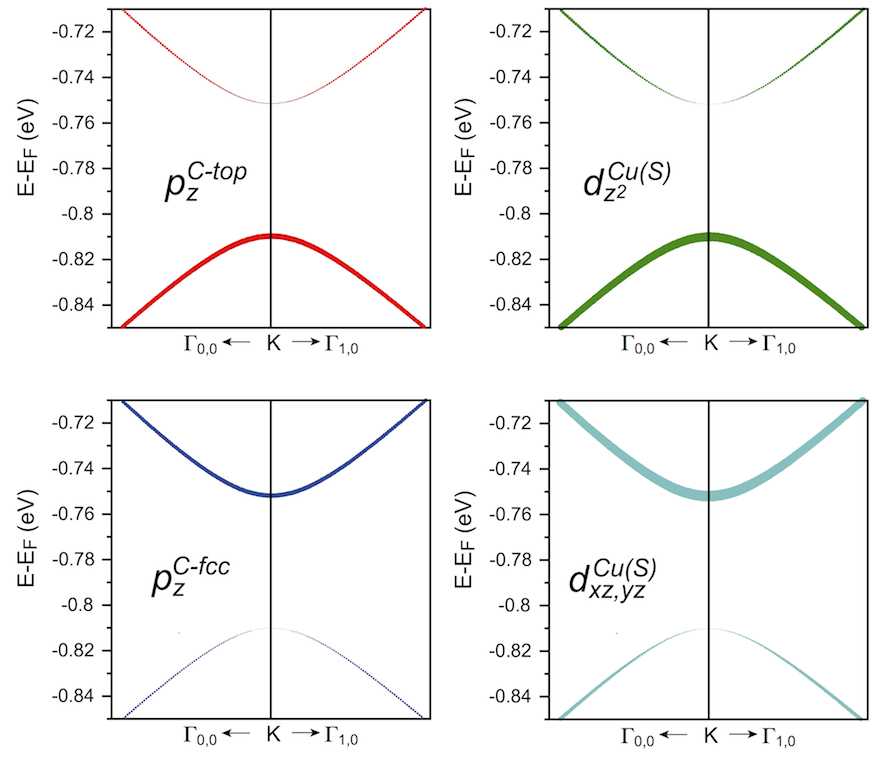}\\
\end{figure}
\noindent\textbf{Fig.\,S4.} Analysis of the electronic structure of graphene/Cu(111) around the $\mathrm{K}$ point corresponding to the graphene-Cu(111) distance of $d=2.0828$\,\AA.

\end{document}